\documentstyle[amsmath,amssymb,graphicx]{aa}

\newcommand{\elie}{{\mathbf E}}

\newcommand{\uin}{u_{\rm in}}

\newcommand{\md}{M_{\rm d}}

\newcommand{\rd}{R_{\rm dg}}
\newcommand{\rp}{R_{\rm pl}}

\newcommand{\rn}{R_{\rm L}}

\newcommand{\re}{R_*}

\newcommand{\ok}{\Omega_{\rm K}}

\newcommand{\cs}{c_{\rm s}}
\newcommand{\dsg}{D_{\rm dg}}

\newcommand{\od}{\Omega_{\rm d}}
\newcommand{\op}{\Omega_{\rm pl}}
\newcommand{\vout}{v_{\rm out}}
\newcommand{\ain}{a_{\rm in}}
\newcommand{\aout}{a_{\rm out}}
\newcommand{\elik}{{\mathbf K}}

\begin{document}

\title{How does disk gravity really influence type-I migration ?}
\author{Arnaud Pierens\inst{1} \and Jean-Marc Hur\'e\inst{1,}\inst{2}\thanks{Now at Universit\'e Bordeaux 1 and Observatoire Astrono\-mique de Bordeaux.}}
\authorrunning{A. Pierens \& J.-M. Hur\'e}
\titlerunning{How does disk gravity really influence type-I migration ?}

\institute{LUTh CNRS UMR 8102 Observatoire de Paris-Meudon, Place Jules Janssen, 92195 Meudon Cedex, France
\and
Universit\'e Paris 7 Denis Diderot, 2 Place Jussieu, F-75251 Paris Cedex 05, France}

\date{Received ??? / Accepted ???}

\abstract{We report an analytical expression for the locations of Lindblad resonances induced by a perturbing protoplanet, including the effect of disk gravity. Inner, outer and differential torques are found to be enhanced compared to situations where a keplerian velocity field for the dynamics of both the disk and the planet is assumed. Inward migration is strongly accelerated when the disk gravity is only accounted for in the planet orbital motion. The addition of disk self-gravity slows down the planet drift but not enough to stop it.}

\maketitle
\section{Introduction}
Interactions between a gaseous disk and embedded proto-planetary cores could be decisive to understand the distribution of semi-major axis, eccentricities and masses of extra-solar giant planets (Perryman 2000, Schneider 2004). Exchanges of angular momentum can efficiently operate through the excitation of spiral density waves at the sites of Lindblad resonances, leading to inward migration of solid bodies (Goldreich \& Tremaine 1979 \& 1980, Ward 1997). For planets with mass less than about a Jovian mass, the interaction between the disk and the planet is mostly linear (type-I migration), in contrast with the non-linear type-II migration where massive planets open a gap. In any case, the drift time-scale is much shorter than the time required for the completion of a (giant) planet. Several works have recently focused on means to slow down or stop migration. The presence of a toroidal magnetic field (Terquem 2003), tri-dimensional effects (Tanaka, Takeuchi \& Ward 2002) or corotation torques (D'Angelo, Henning \& Kley 2002) could act in this way.

From numerical simulations, Nelson \& Benz (2003a \& b) have shown that disk self-gravity can noticeably affect the drift velocity (even for low mass disks). They suggested that even very weak changes of the rotation curve induced by disk gravity significantly modifiy the location of Lindblad resonances, and subsequently the total differential torque. 
Their conclusions are however strongly resolution-dependent. In this short communication, we clarify the influence of disk gravity on type-I migration by a semi-analytical approach. In particular, we determine analytically for the first time the location of Lindblad resonances modified by disk gravity, and compute the corresponding gravitational torques. Analytical techniques generally provide reliable diagnostic tools and powerful predictions as they implicitely correspond to an infinite numerical resolution.
 
 In Sect. \ref{sec:location}, we derive and discuss a general expression for the location of Lindblad resonances as functions of the disk surface density profile, orbit of the planet, relative disk mass and edges. In Sect. \ref{sec:influence}, we successfully check this expression in the simple case of radially homogenous disks. The effect of the disk mass on Lindblad torques is then analyzed. Finally, we consider the case of disks with power-law surface density profiles. We conclude in Sect. \ref{sec:conc}.

\section{Location of Lindblad resonances}
\label{sec:location}

\subsection{Background}

A planet embedded in a gaseous disk exerts gravitational torques at sites of Lindblad resonances. For low mass bodies, the disk response to the perturbing point mass potential can be considered as linear, so that torques can explicitly be determined (Goldreich \& Tremaine 1979, Artymowicz 1993). The nominal positions $\rn(m)$ of the inner (ILRs) and outer (OLRs) Lindblad resonances associated with the $m$-th order Fourier pattern are found from the equation $D(\rn)=0$ with
\begin{equation}
D = \kappa^2-m^2(\Omega -\op)^2,
\label{eq:d}
\end{equation}
where $\kappa$ is the epicyclic frequency defined by $\kappa^2=4\Omega^2+2R\Omega d\Omega/dR$, $\Omega$ is the fluid angular velocity and $\op$ is the angular velocity of the planet. Generally,  $\rn(m)$ differs from the effective locations of  Lindblad resonances  (hereafter $\re$) where the waves become evanescent.  Effective resonances are found from the equation $D_*(\re)=0$ with (Artymowicz 1993, Ward 1997)
\begin{equation}
D_* = D +\frac{m^2\cs^2}{R^2}, 
\label{eq:dstar}
\end{equation}
where $\cs$ is the sound speed. If the disk aspect ratio $H/r \equiv \eta$ is a constant, we then have $\cs = \ok \eta R$, and so
\begin{equation}
\re=\rp \left(1+\epsilon \frac{f}{m}\right)^{2/3},
\label{eq:reff}
\end{equation}
where $\epsilon=-1$ stands for the ILRs, $\epsilon=+1$ is for the OLRs, and $f=\sqrt{1+m^2\eta^2}$. This expression shows that nominal and effective resonances coincide only for low $m$ values.

\subsection{Effects of disk gravity}

The gas in the disk, whatever its mass relative to the central mass, is a source of gravity. It modifies the positions of the Lindblad resonances in two ways. First, it changes the dynamics of both the gas component itself (i.e. disk self-gravity) and the planet according to
\begin{equation}
\begin{cases}
\Omega^2 = \ok^2+\od^2,\\
{\op'}^2= \op^2+\omega^2,\\
\end{cases}
\end{equation}
where $\od^2$ is the contribution due to the disk, and $\omega^2 = \od^2(\rp)$. Second, it shifts the location of effective resonances closer to the planet (density waves can propagate between Lindblad resonances). According to Nelson \& Benz (2003b), the new locations $\rd$ of effective ILRs and OLRs modified by the disk gravity are determined by the equation $\dsg(\rd)=0$ where
\begin{equation}
\dsg = D_* -\frac{2\pi G\Sigma m}{R},
\label{eq:dsg}
\end{equation}
and $\Sigma$ is the disk local surface density.

\subsection{Second-order differential method}

Provided the relative shift $\delta R \equiv \rd - \re$ is small compared to $\re$, we can derive a reliable analytical expression for $\rd$ by a low order expansion of $\dsg(\re + \delta R)$. Keeping terms of second order in $\od/\ok$ only, we find after some algebra
\begin{equation}
\label{eq:totalshift}
\delta R  =  \delta R_1+ \delta R_2 + \delta R_3
\end{equation}
where
\begin{equation}
\begin{cases}
\frac{\delta R_1}{\re}  =  \frac{\od^2 \xi }{3\ok^2}\left( 4 +  \gamma + m \epsilon f\right)\\
 \frac{\delta R_2}{\re} = - \frac{2 \pi G m \Sigma  \xi }{3 R \ok^2}\\
 \frac{\delta R_3}{\re} =  - \frac{\omega^2 \xi f m^2}{3 \ok^2 (\epsilon m + f)},
\label{eq:shifts}
\end{cases}
\end{equation}
with
\begin{equation}
\begin{cases}
\gamma &= \frac{d \ln |\od^2|}{d \ln R},\\
\xi &= \frac{1}{1+m \epsilon f + m^2 \eta^2}.
\end{cases}
\end{equation}

All quantities in Eq. (\ref{eq:shifts}) are to be expressed at $\re$. The first two terms in the right-hand side of Eq.(\ref{eq:totalshift}) are due to the disk self-gravity, whereas the third one corresponds to the change of the planet angular velocity due to the disk mass.

\begin{table}
\begin{center}
\begin{tabular}{lcccc}
 shift             & ILRs    &  OLRs     & contribution & inward migration\\ \hline 
$\delta R_1$                 &  $>0$     & $>0$                 & outwards & slowed down \\ 
$\delta R_2 $      & $>0$    &  $<0$     & inwards & accelerated\\ 
$\delta R_3$       & $<0$    &  $<0$     & inwards & accelerated \\ 
total & $>0$    &  $<0$     & {\bf inwards} & {\bf accelerated} \\  \hline
\end{tabular}
\end{center}
\caption{Expected sign of the three individual shifts, and corresponding prediction about planetary migration.}
\label{tab:droverr}
\end{table}

The ``efficiency'' of migration can qualitatively be deduced from the shifts $\delta R/\re$ of the ILRs and OLRs since, in general, the amplitude of gravitational torques depends strongly on the location of the resonances. Further, these torques are the largest for intermediate $m$ values $\sim 10-20$ ($\sim 1/\eta$). In these conditions, we have $f \sim 1$, $\xi \sim 1/m \epsilon$. Using the (crude) monopole approximation for $\od^2$ (Mineshige \& Umemura 1996) and assuming a power law surface density profile, we find
\begin{equation}
\label{eq:simple}
\begin{pmatrix}
\delta R_1\\
\delta R_2\\
\delta R_3
\end{pmatrix}
\sim \re \times \frac{\od^2}{3\ok^2}
\begin{pmatrix}
1+\frac{(4+\gamma)\epsilon}{m}\\
- \epsilon \\ 
- 1+ \frac{\epsilon}{m}\\
\end{pmatrix}.
\end{equation}

Note that $|\gamma|$ is of the order of unity in most cases. To conclude on the influence of disk gravity, one must compare each term in Eq.(\ref{eq:simple}) together. Table \ref{tab:droverr} summarizes the sign of each shift, and the consequence on migration. We see that if self-gravity is neglected (i.e. assuming $\delta R\approx \delta R_3$), then inward migration is accelerated. This effect has been clearly observed in the hydrodynamical simulations by Nelson \& Benz (2003a \& b). On the contrary, self-gravity alone (i.e. $\delta R \approx \delta R_1 + \delta R_2$) slows down migration due to an asymetrical shift of the resonances. Further, since
\begin{equation}
\left| \frac{\delta R_1 + \delta R_3}{\delta R_2} \right| \sim \frac{(5+\gamma)}{m} < 1,
\end{equation}
we conclude that the total shift is dominated by the term $\delta R_2$. From Eq.(\ref{eq:simple}), we have
\begin{flalign}
\left.\frac{\delta R}{R}\right|_{\rm OLR} + \left.\frac{\delta R}{R}\right|_{\rm ILR} & \sim - \left(1-\frac{5+\gamma}{m}\right)  \\
 & \qquad  \qquad \times \frac{4 \rp}{9 m} \frac{d}{dR} \left( \frac{\od^2}{\ok^2} \right) < 0
\nonumber
\end{flalign}
For most disks of astrophysical interest, we expect that $\od^2/\ok^2$ to be an increasing function of the radius. It then follows that OLRs should be shifted more than ILRs, meaning that {\it migration tends to proceed inward more rapidly when disk gravity in included}.

\section{Influence of the disk gravity on low mass planet migration}
\label{sec:influence}

\subsection{Gravitational acceleration in an homogeneous disk}

As Eq.(\ref{eq:shifts}) shows, resonance shifts due to the disk mass can be determined once the radial gravity field $g_R = - \od^2 R$ is known. Since there is no reliable formula for potential/density pairs in flat disks, we shall consider a simple case which allows some analytics, namely a thin disk with uniform surface density $\Sigma$ (a case often considered in simulations). Then, the radial field $g_R$ inside the disk exactly writes (e.g. Durand 1964)
\begin{equation}
g_R=4G\Sigma\left[\frac{\elie(\vout)-\elik(\vout)}{\vout}+\elik(\uin)-\elie(\uin)\right]
\label{eq:gr}
\end{equation}
where $\elik$ and $\elie$ are the complete elliptic integrals of the first and second kinds respectively, $\uin =\ain/R \le 1$, $\vout = R/\aout \le 1$, where $\ain \ge 0$ is the disk inner edge and $\aout > \ain$ is the outer edge. A more tractable expressions for $g_R$ can be derived from truncated expansions of the complete elliptic integrals. For instance, with a classical second-order expansions over the modulus $x < 1$ (e.g. Gradshteyn \& Ryzhik 1994), namely
\begin{equation}
\begin{cases}
\elik(x)&= \frac{\pi}{2} \left( 1 + \frac{1}{4}x^2  \right)  + {\cal O}(x^4),\\
\elie(x)&= \frac{\pi}{2} \left( 1 - \frac{1}{4}x^2  \right)  + {\cal O}(x^4),
\end{cases}
\label{eq:kseries}
\end{equation}
the field far from edges is given in a good approximation by
\begin{equation}
g_R \approx -\pi G \Sigma \left(\vout - \uin^2 \right). 
\label{eq:grapprox}
\end{equation}
Note that contrary to Eq.(\ref{eq:gr}), that formula remains finite at the edges, and then is more realistic although approximated. Since the gravitational potential of the disk is minimum very close to the inner edge (like in most astrophysical disks), we can simplify Eq.(\ref{eq:grapprox}) by considering only the term linear in $R$, and so
\begin{equation}
\od^2 \sim \frac{\pi G \Sigma}{\aout}.
\end{equation}

\subsection{Shifts and torques at the Lindblad resonances}

\begin{figure}
\includegraphics[width=8.75cm]{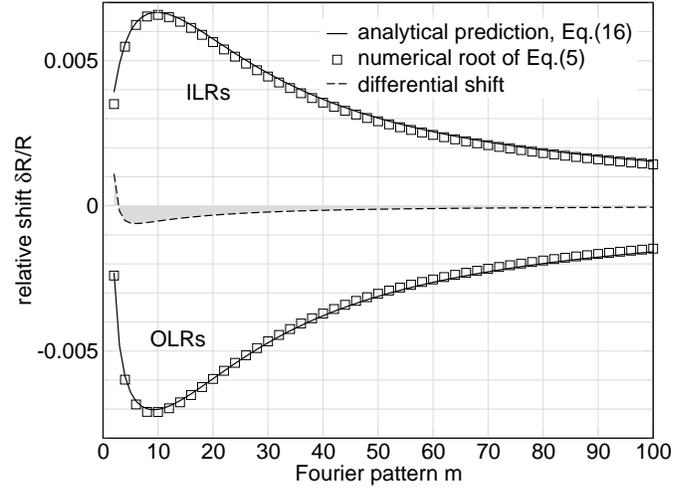}
\caption{Relative shift  $\delta R/\re$ of the effective ILRs and OLRs due to the disk mass, for the following parameters: $\ain=0.01$, $\aout=1$, $\mu=0.05$ and $\rp/\aout=~0.5$.}
\label{fig:anavsnum.eps}
\end{figure}

\begin{figure}
\includegraphics[width=8.7cm]{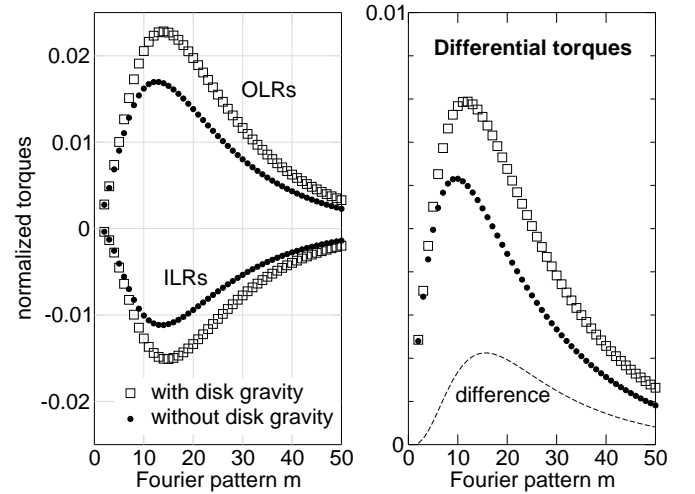}
\caption{Inner and outer Lindblad torques ({\it left}) and differential torques ({\it right}) when the disk mass is accounted for, compared to the case without disk gravity.  The conditions are the same as for Fig. \ref{fig:anavsnum.eps}.}
\label{fig:torques.eps}
\end{figure}

In the homogeneous disk model, we thus have
\begin{equation}
\frac{\od^2}{\ok^2} \sim  \frac{\mu R^3}{\aout^3},
\end{equation}
where $\mu=\md/M$, $\md$ being the disk mass, and assuming $\ain^2 \ll \aout^2$. Hence, $\gamma=0$. Using Eq.(\ref{eq:reff}), our formula for the relative shift becomes
\begin{equation}
\frac{\delta R}{\re} =  \frac{\mu \rp^3}{3\aout^3} \left(\frac{m+\epsilon f}{m}\right)^2\left(\frac{ 4 +  \frac{mf^2}{m+\epsilon f}- 2m \frac{\aout}{\re}}{ 1+m \epsilon f + m^2 \eta^2 }\right).
\label{eq:shifthomo}
\end{equation}
We see that the relative shift is strongly sensitive to the location of the planet with respect to the disk outer edge. Figure \ref{fig:anavsnum.eps} shows $\delta R/\re$ as predicted by Eq.(\ref{eq:shifthomo}) for typical parameters. The agreement between the above formula and values determined numerically from Eq.(\ref{eq:dsg}) by standard root finding techniques is excellent. Both Lindblad resonances get closer to the planet. The OLRs are slightly more shifted than the ILRs, specially for small $m$ (depending on the disk aspect ratio). This can be understood by computing the relative displacement
\begin{equation}
\left.\frac{\delta R}{R}\right|_{\rm OLR} + \left.\frac{\delta R}{R}\right|_{\rm ILR} \approx \frac{8\mu \rp^3}{3m^2\aout^3} \left(1 - \frac{m \aout}{2 \re} \right) \lesssim 0.
\label{eq:shiftdiff}
\end{equation}
 Figure \ref{fig:torques.eps} displays the Lindblad torques computed following the procedure described in Ward (1997). It confirms that the outer torques have larger amplitude than inner ones when the disk gravity is accounted for. Differential torques are larger, and inwards migration should be accelerated.

\begin{figure}
\includegraphics[width=8.8cm]{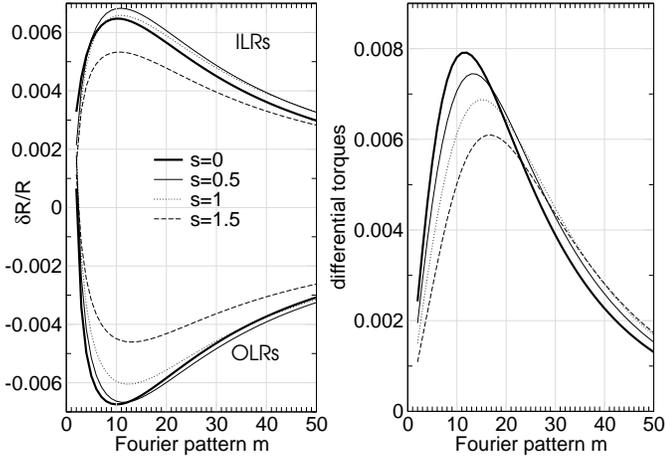}
\caption{Relative shifts of the OLRs and ILRs ({\it left}) and differential Lindblad ({\it right}) for different power law surface density profiles (see text). The conditions are the same as for Fig. \ref{fig:anavsnum.eps}.}
\label{fig:vss.xfig.eps}
\end{figure}

\subsection{The case of non homogeneous disks}

We have computed the resonance shifts and associated Lindblad torques for disks with power law surface density profiles (i.e. $\Sigma \propto R^{-s}$), typical from disk models and observations. The disk gravity field $g_R$ has been determined numerically from the splitting method described in Hur\'e \& Pierens (2004). As shown in Fig. \ref{fig:vss.xfig.eps}, resonances shifts are weakly affected by the surface density profile for $s=1/2$ and $1$. For steeper profiles however (like for $s=3/2$), shifts are smaller. Figure \ref{fig:vss.xfig.eps} also displays the differential torques. These conserve the same shape as in the homogeneous case, are always larger than in the case without disk gravity. Their magnitude decreases as the surface density profile gets steeper, especially for low $m$ values. Changes are minor for high Fourier modes. Globally, the conclusions established in the homogeneous disk model still hold: inward migration should proceed faster due to the disk gravity. For large $s$-values, the effect of disk gravity on migration is predicted to be less and less efficient.

\begin{figure}[h]
\includegraphics[width=8.7cm]{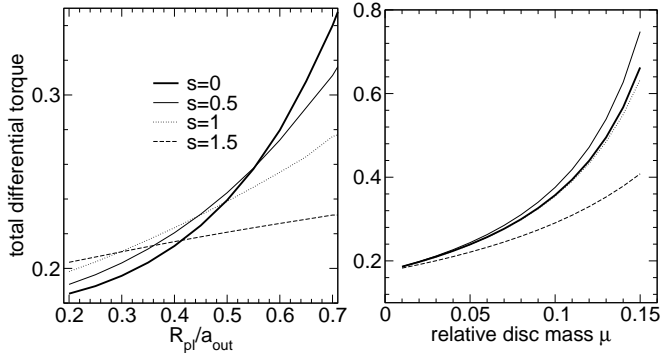}
\caption{Total differential torque versus the planet position relative to the outer edge for $\mu=0.05$ ({\it left}), and versus the relative disk mass $\mu$ for $\rp=0.5$ ({\it right}), for different power law surface density profile in the disk (see text).}
\label{fig:mig.xfig.eps}
\end{figure}

Figure \ref{fig:mig.xfig.eps} displays the differential torque as functions of $\rp$ and $\mu$ for various exponents $s$ in the power law surface density. We see that torques increase as the planet orbits closer to the outer edge, and as the disk mass rises. Both effects are due to the fact that the the resonances get closer to the planet as $\mu$ or $\rp/\aout$ increases, however with a slightly larger shift of the OLRs with respect to the ILRs (for reasons explained above).
 
\section{Concluding remarks}
\label{sec:conc}

In this paper, we have reported a general expression for the shift of the Lindblad resonances due to the disk gravity, whatever the surface density profile, and computed associated torques. In contrast with current numerical simulations (Nelson \& Benz 2003a \& b), our analysis is not resolution-dependent, thereby enabling reliable predictions about the migration mechanism of low mass embedded objects. We have considered the effect of the disk gravity i) on the planet dynamics, and ii) on the disk itself (i.e. self-gravity). Both effects are important and act in opposite ways. We confirm that disk gravity plays an important role on type-I migration, even for low mass disks. We find that the position of the resonances are significantly modified and get closer to the planet when the disk mass is taken into account. The differential Lindblad torques are stronger than in the case where the disk mass is neglected (Ward 1997). Our results are also compatible with the recent simulations by Nelson \& Benz (2003a \& b): migration is accelerated when the disk gravity is accounted for in the motion of the planet only, and slowed down when self-gravity is added, but does not stop it (assuming that all Fourier modes exist).  Regarding extrasolar planets, our conclusions reinforce the necessity to seek for mecanisms able to cancel inward migration. We note that the possible suppression of low $m$-modes of the OLRs (for instance if the planet evolves too close to the disk outer edge) could decrease the total torque exerted on the planet and change its drift.

It would be of great interest i) to seek for a general expression of $\delta R$ as a function of the surface density profile (for instance as an explicit function of the $s$-exponent), and ii) to compare the location of Lindblad resonances and associated torques as predicted here with those obtained directly by numerical simulations of fully self-gravitating disks. This would require very high numerical resolutions.


\begin{thebibliography}{}

\bibitem[Artymowicz(1993)]{1993ApJ...419..155A} Artymowicz, P.\ 1993, ApJ, 
419, 155 

\bibitem[D'Angelo, Henning, \& Kley(2002)]{2002A&A...385..647D} D'Angelo, 
G., Henning, T., \& Kley, W.\ 2002, A\&A, 385, 647 

\bibitem{} Durand E., 1964, in {\it ``Electrostatique''}, Masson Ed.

\bibitem[\protect\citeauthoryear{Goldreich \& Tremaine}{1979}]{1979ApJ...233..857G} Goldreich P., Tremaine S., 1979, ApJ, 233, 857
 
\bibitem[Goldreich \& Tremaine(1980)]{1980ApJ...241..425G} Goldreich, P.~\& 
Tremaine, S.\ 1980, ApJ, 241, 425 

\bibitem{} Gradshteyn I.S., Ryzhik I.M., 1994, in {\it ``Table of integrals, series and products''}, Academic Press Inc., 5th ed., A. Jeffrey Ed.

\bibitem{} Hur\'e J.M. \& Pierens A., 2004, ApJ, in press

\bibitem[Mineshige \& Umemura(1996)]{1996ApJ...469L..49M} Mineshige, S., \& 
Umemura, M.\ 1996, ApJL, 469, L49 

\bibitem[Nelson \& Benz(2003)]{2003ApJ...589..556N} Nelson, A.~F.~\& Benz, 
W.\ 2003a, ApJ, 589, 556 

\bibitem[Nelson \& Benz(2003)]{2003ApJ...589..578N} Nelson, A.~F.~\& Benz, 
W.\ 2003b, ApJ, 589, 578 

\bibitem[\protect\citeauthoryear{Perryman}{2000}]{2000RPPh...63.1209P} Perryman M.~A.~C., 2000, RPPh, 63, 1209 

\bibitem[]{} Schneider J., 2004, ``The Extrasolar Planets Encyclopaedia'', http://www.obspm.fr/planets 

\bibitem[\protect\citeauthoryear{Tanaka, Takeuchi, \& Ward}{2002}]{2002ApJ...565.1257T} Tanaka H., Takeuchi T., \& Ward W.~R., 2002, ApJ, 565, 1257 

\bibitem[\protect\citeauthoryear{Terquem}{2003}]{2003MNRAS.341.1157T} Terquem C.~E.~J.~M.~L.~J., 2003, MNRAS, 341, 1157

\bibitem[Ward(1997)]{1997Icar..126..261W} Ward, W.~R.\ 1997, Icarus, 126, 261 

\end{thebibliography}
\end{document}